\documentclass[russian,english,english]{article}
\usepackage[T1]{fontenc}
\usepackage[koi8-r,latin9]{inputenc}
\usepackage{esint}

\makeatletter
\@ifundefined{date}{}{\date{}}
\newtheorem{theorem}{Theorem}

\newtheorem{lemma}[theorem]{Lemma}

\usepackage{babel}

\makeatother

\usepackage{babel}
\begin{document}

\title{Mathematics for some classes of networks}

\author{V. A. Malyshev, A. A. Zamyatin}

\maketitle
\begin{center}

\noindent Lomonosov Moscow State University, Moscow, Russia 

\noindent E-mail addresses: malyshev2@yahoo.com, andrew.zamyatin@gmail.com

\end{center}

\begin{abstract}

Network (as a general notion) is not a mathematical object - there
is no even any definition. However, there is a lot of good rigorous
mathematics for well-defined classes of networks. In sections 1-3
we give a short overview of classes of networks which interested the
authors for some time. In section 4 we consider in detail a new class
of networks, related to markets with many agents.

\end{abstract}

\section{Random field dynamics on a fixed graph}

The basic element of most networks is a graph $G$ with the set $V=V(G)$
of vertices and the set $L=L(G)$ of links (lines, edges). Second
basic element is a function $s=f(v):V\to S$ with values in some space
$S$ The elements of $S$ may be called marks, spins, field values,
queues etc. The function $f$ is subjected to random dynamics.

Simplest example is (an earlier stuff) random walks on graph, where
$f=0$ everywhere except one point where the particle is situated.
This is related to electric networks, see for example \cite{bo}.

In general there are two different situations. First one is a local
continuous time Markov dynamics given by infinitesimal transitions.
Classical reference is \cite{Liggett}, mostly such processes model
stochastic dynamics of particles or spins. The latter are related
to Gibbs random fields (invariant measures for this dynamics) on graphs,
see \cite{m4} and references therein.

\paragraph{Queuing, communication and transportation networks }

The simplest case is when the particles jump (from one node to another)
freely without seeing each other, the only interaction is only through
queues at the nodes, where they spend some time. There are two main
theories concerning such class of networks: 
\begin{enumerate}
\item Most popular - Jackson network (1963) and its generalizations (Gordon-Newel,
BCMP). This theory gives explicit formulas for the stationary distribution
and is the origin of many other analytical results. One of the applications
is to describe jams and phase transitions in communication \cite{m5}
and transportation \cite{m6} networks. 
\item Stability theory (1968-1995) exhibits in many cases of large time
qualitative behavior. If the walking clients are identical then it
is described by random walks in orthants and strongly uses It uses
Lyapounov functions, Euler scaling (fluid approximation), ergodic
theory of dynamical systems and Lyapounov exponents, see \cite{m7}.
If the walking clients can be of finite number of types then the corresponding
theory \cite{m3} is the union of the one type case and the theory
of random grammars (see below). 
\end{enumerate}
For more sophisticated restrictions - network protocols (TCP etc.)
- there are many partial results but no comparable (deeply elaborated)
mathematical theories.

\textbf{Chemical kinetics - mean field Markov chain}

Mean field network means that there is no specified local structure
on the graph. Example of such theory is the chemical kinetics. It
describes the following situation. Molecular types are indexed by
$V$, $n_{v}$ - number of molecules of type $v$ 
\[
n_{1}+...+n_{|V|}=N
\]
There are also reaction types $r=1,2,...,R$, formally - multigraph
defined by finite number of equations 
\[
\sum_{v}s_{vr}M_{v}=0
\]
where $M_{v}$ - molecule of type $v$, $s_{vr}$ - stoihiometric
coefficients of molecule type $v$ in reaction of type $r$ , negative
for substrates, positive for products. Reaction rates (continuous
time Markov chain) are given by

\[
\lambda_{r}=A_{r}\prod_{v:s_{vr}<0}n^{-s_{vr}}
\]
for the jump (transition) 
\[
n_{v}\to n_{v}+s_{vr},v\in V
\]
To get ODE of \textbf{classical chemical kinetics} 
\[
\frac{dc_{v}}{dt}=\sum_{r}Q_{vr}(c_{1},...,c_{|V|})
\]
for some polynomials $Q_{r}$, in the limit $N\to\infty$ 
\[
c_{v}(t)=\lim\frac{n_{v}^{(N)}(t)}{N}
\]
one uses canonical scaling of reaction rates 
\[
A_{r}=a_{r}N^{s_{r}+1},s_{r}=\sum_{v:s_{vr}<0}s_{vr}
\]
To deduce \textbf{chemical thermodynamics} is more difficult \cite{m8},
one should, together with molecular types $v$, introduce more degrees
of freedom: kinetic energy $T_{v,i}$ and internal energy $K_{v.i}$
of $i$-th molecule of type $v$. Also, one should define more complicated
mean field dynamics - introduce energy mechanism in reactions. As
there is kinetic energy - there should be Newtonian movement, and
the dynamics become mixed: \textbf{local + mean field}. Molecule move
freely (as in ideal gas) but kinetic energies randomly interchange
with internal energies.

\paragraph{Network homeostasis \cite{m9}}

Network is defined by 
\begin{enumerate}
\item large graph $G$ of compartments, this graph $G$ has metrics and
the boundary, 
\item in any compartment chemical kinetics is defined, that is there are
molecules with chemical reactions, 
\item there is transport of molecules between compartments 
\item there is input and output of molecules on the boundary 
\end{enumerate}
Under some conditions (the main is that reactions are unary) it is
possible to prove that \textbf{far from the boundary there is equilibrium}
- concentrations almost do not change with change of input.

\section{Dynamics of graphs and of marked graph}

Earlier the science of random graphs considered mainly the properties
of graphs with fixed number of vertices and/or random number (for
example Bernoulli) of edges, see for example \cite{bo,Janson,Kolchin,kar}.
The simplest dynamics (appending edge by edge) appeared already in
\cite{erre}, see also \cite{pal}. What more general dynamics on
graphs one should study? First of all, it is more reasonable to consider
evolution of marked graphs. Most general dynamics of marked graphs
(local random dynamics of a graph. jointly with a field on it) is
called random graph grammars \cite{hab,m1,ros} and \cite{m12,m13}.
It appears to be quite natural in connection with the emerging new
physical theories \cite{m10,m11,m17} and social networks \cite{Sattoras}.
Namely, if eventually the local space-time appears to be discrete,
then the most natural language for it is a graph with some physical
fields on it. The dynamics of the space time is local. The example
is the following.

\paragraph{Macrodimension of a graph - invariant of local dynamics}

We consider infinite (countable) graphs $G$. Let $O_{n}(v)$ be the
neighborhood of vertex $v$ of radius $n$. Put 
\[
D_{n}(v)=\frac{\ln\left|O_{n}(v)\right|}{\ln n},\overline{D}(v)=\lim\sup_{n\rightarrow\infty}D_{n}(v),\underline{D}(v)=\lim\inf_{n\rightarrow\infty}D_{n}(v)
\]
If for all $v$ 
\[
\overline{D}(v)=\underline{D}(v)=D_{S}
\]
then $D_{S}$ is called \textbf{scaling macrodimension} of graph $G$.
For example any homogeneous lattice in euclidean space $R^{d}$ has
scaling macrodimension $D_{S}=d$. Note that there are many other
definitions of variants of macrodimension: connectivity, Hausdorf,
entropy, inductive macrodimension.

Denote $\mathbf{G}_{M}$ the class of connected graphs where each
vertex has degree $\leq M$. Let $U$ - \textbf{any local dynamics}
(graph grammar).

There is the following result \cite{m10}. If for some sufficiently
large $M$ $U$ leaves the \textbf{class $\mathbf{G}_{M}$ invariant}
and the corresponding Markov chain is \textbf{locally reversible}
then the scaling macrodimension is an invariant.

Local reversibility means that Kolmogorov cycle criteria relations
\[
a_{i_{1}i_{2}}...a_{i_{L}i_{1}}=1,a_{ij}=\frac{\lambda_{ij}}{\lambda_{ji}}
\]
follow from such relations of bounded length.

\paragraph{Random graph grammars }

Consider words $\alpha=x_{1}...x_{N}$ (ordered sequences of symbols),
where $x_{N}$ belongs to some finite alphabet $A$. Grammar is defined
by the list $Sub$ of productions (allowed substitution types) 
\[
S_{j}:\,\,\,\,\alpha_{j}\to\beta_{j},j=1,...,S
\]
Random grammar includes also positive numbers $\lambda_{j}$ (rates).
That is at time interval $(t,t+dt)$ in the word $\alpha(t)$ any
subword $\alpha_{j}$ is independently replaced by $\beta_{j}$ with
probability $\lambda_{j}dt$ (continuous time Markov chain).

For \textbf{graph grammar} $\alpha(t)$ are marked graphs, $\alpha_{j},\beta_{j}$
are (small) connected marked graphs. Thus, $\alpha_{j}$ is deleted
from the graph and $\beta_{j}$ is pasted instead (some restrictions
needed of course). Note that ordinary grammar is a particular case,
corresponding to linear marked graphs.

One of the problems - invariant measure and conserved characteristics
with respect to given graph grammar dynamics was considered in \cite{m6,m12,m13}.

\section{Quantum Graph Grammar}

What is quantum graph \cite{m18}. Consider Hilbert space $l_{2}(\{G\})$
with (orthonormal) basis $e_{G}$, enumerated by finite graphs $G.$
Or by finite marked graphs if the set of marks is finite.\textbf{
}First example is linear marked graphs - \textbf{quantum words}.

To define quantum dynamics assume that if $a_{j}=(\alpha_{i}\to\beta_{i})\in Sub$
then also inverse substitution $a_{j}^{*}=(\beta_{i}\to\alpha_{i})\in Sub$.
Denote $S_{j}(k)$ the substitution $S_{j}$ applied to subword of
the word $\alpha$ starting on $k$-th symbol of the word $\alpha$.
Introduce the Hamiltonian 
\[
\sum_{j=1}^{|Sub|}\sum_{k=1}^{\infty}(\lambda_{j}a_{j}(k)+\lambda_{j}^{*}a_{j}^{*}(k))
\]
The first simple result is: this Hamiltonian is selfadjoint in $l_{2}(\{G\})$,
that is the quantum evolution is well-defined.

\paragraph{Gibbs and Quantum Spaces}

$\mathbf{G}$ - class of finite graphs with a function $f:V\to R$
, called spin graphs $(G,f)$, $\mathbf{G}_{N}$ - class of such spin
graphs of radius $\leq D$. Potential is defined as some function
$\Phi:\mathbf{G}_{D}\to R$. Hamiltonian $H:\mathbf{G}\to R$ is 
\[
H((G,f))=\sum\Phi(\gamma)
\]
where the sum over all sub spin subgraphs of $(G,f)$. Partition function
\[
Z_{N}=\sum_{(G,f)\in G_{N}}\exp(-\beta H(G,f))
\]
Gibbs measure on $\mathbf{G}_{f,N}$ 
\[
\mu_{N}(G,f)=Z_{N}^{-1}\exp(-\beta H(G,f))
\]

There are many results-examples (by physicists and mathematicians)
related to ``quantum gravity'', see for example \cite{m11,m17}
and references therein.

\section{Trading network as Boltzmann mechanics of communicating vessels}

Standard financial mathematics considers games of one or small number
of players against the chance (random market). Recently, a new approach
(called multi-agent models) appeared which considers the games of
many players against each other. This theory is at the starting point
and its models are mainly \textbf{mean-field} models.

In this section some \textbf{local} models are considered where there
are many players and many financial or trading instruments. Our model
develops simpler models of (\cite{m15,m16}). The model resembles
communication and transportation networks - the main difference is
that the nodes have special dynamical values (moving boundaries, or
real prices). The clients have also their own subjective prices and
their interaction (transaction) with the nodes depend on these prices.
This model does not describe any real situation (and any other existing
multi-agent model as well) but we hope that some features of this
model will be useful for future more realistic models.

\paragraph{Free\foreignlanguage{russian}{ one-phase Boltzmann dynamics }}

Consider the phase space $\mathbf{S}=I\times I_{0}$, where $I\subset R$
is an infinite interval and $I_{0}=[-V_{0},V_{0}],\,0<V_{0}<\infty$.
On $\mathbf{S}$ at any time $t\geq0$ a random locally finite configuration
$\{(x_{i}(t),v_{i}(t))\}$ of particles is given with coordinates
$x_{i}\in I$ and velocities $v_{i}\in I_{0}$. Assume that this configuration
at any time $t$ has distribution $P_{t}$ with one-particle correlation
function $f(x,v,t)$ defined so that for any subset $A\subset\mathbf{S}$
of the phase space 
\[
E\,\#\{i:\,(x_{i},v_{i})\in A\}=\int_{A}f(x,v,t)dxdv
\]
One can have in mind Poisson measure $P_{0}$ at time $t=0$. Any
particle moves always with its initial velocity, independently of
other particles. Also there is Poisson income flow of particles from
exterior with rate $\lambda(x,v,t)$, that is during time interval
$[t,t+dt]$ the mean number of incoming particles to the cell $[x,x+dx]\times[v.v+dv]$
of the phase space is $\lambda(x,v,t)dxdvdt$. Assume moreover that
each particle can die (disappear) with exponential distribution having
rate $\mu(x,v,t)$. This means that during time $dt$ $\mu(x,v,t)dxdvdt$
particles leave the cell $dxdv$.

Remind that we assume boundedness of velocities, that is 
\[
f(x,v,t)=\lambda(x,v,t)=\mu(x,v,t)=0,|v|\geq V_{0}
\]
\begin{lemma}

For any $x\in I$ and $t<\frac{d(x,\partial I)}{V_{0}}$, where $d(x,\partial I)$
is the distance of the point $x$ from the boundary of $I$, the standard
linear Boltzmann equation holds 
\begin{equation}
\frac{\partial f}{\partial t}+v\frac{\partial f}{\partial x}=-\mu(x,v,t)f(x,v,t)+\lambda(x,v,t)\label{free_Boltzman}
\end{equation}

\end{lemma}

This is trivial for $\mu=\lambda=0$. In fact, for small $\delta>0$
we have 
\begin{equation}
f(x,v,t+\delta)=f(x-v\delta,v,t)\label{0approx}
\end{equation}
if $x$ is not on the boundary of $I$ and $\delta$ is sufficiently
small. Subtracting $f(x,v,t)$ from both parts of this equality, dividing
by $\delta$ and taking the limit $\delta\rightarrow0$, we have 
\begin{equation}
\frac{\partial f}{\partial t}+v\frac{\partial f}{\partial x}=0\label{kin}
\end{equation}
The unique solution of the Cauchy problem for (\ref{kin}) is 
\[
f(x,v,t)=f(x-vt,v,0)
\]
If $\lambda\neq0,\mu=\mu(x,v)\neq0$ then it is also easy to see that
the equation (\ref{free_Boltzman}) holds. Note that if $\lambda=0$
and $\mu$ does not depend on $t$, there is also explicit solution,
see section XI.12 in \cite{ReedSimon} 
\[
f(x,v,t)=f(x-vt,v,0)\exp(-\int_{0}^{t}\mu(x-vs,v)ds)
\]

\paragraph{Two phases - particle dynamics}

We shall define two types of dynamics - particle dynamics and continuum
media dynamics.

In the particle dynamics $(\pm)$-phases consist of $(\pm)$-particles
so that each $(-)$-particle is to the left of any $(+)$-particle.
Denote $b(t)\in R$ (boundary between phases) the coordinate of the
leftmost $(+)$-particle. Then for $x\geq b(t)$ there is $(+)$-phase
and for $x<b(t)$ there is $(-)$-phase. Particles move, as above,
with their own velocities until a $(-)$-minus particle reaches the
point $b(t)$, then it disappears together with the $(+)$-particle
at $b(t)$ and the point $b(t)$ jumps to the coordinate of the new
leftmost $(+)$-particle. After this, the process proceeds similarly.

Random configurations of particles are defined by the correlation
functions $f_{\pm}(x,v,t)$ correspondingly. Assume that also the
functions $\lambda_{\pm}(r,t),\mu_{\pm}(r,t),r\geq0,$ are defined,
smooth on $R_{+}$ and zero if $r\geq R_{0}$ for some $0<R_{0}<\infty$.

The dynamics of one point correlation functions \inputencoding{koi8-r}\foreignlanguage{russian}{$f_{\pm}(x,v,t)$
}\inputencoding{latin9}for $x\neq b(t)$,\inputencoding{koi8-r}\foreignlanguage{russian}{
that is on $(b(t),\infty)$ and $(-\infty,b(t))$ correspondingly}\inputencoding{latin9},
\inputencoding{koi8-r}\foreignlanguage{russian}{is given by the equations
(already non-linear as $b(t)$ is unknown) 
\begin{equation}
\frac{\partial f_{\pm}}{\partial t}+v\frac{\partial f_{\pm}}{\partial x}=-\mu_{\pm}(x-b(t),v,t)f_{\pm}(x,v,t)+\lambda_{\pm}(x-b(t),v,t)\label{Boltzman}
\end{equation}
This means that we assume that arrivals and departures depend only
on the distance $r=|x-b(t)|$.}

\inputencoding{latin9}Thus two phases add reactions between particles
of different phases. The following interpretation is useful. We consider
one instrument (\textbf{stocks, futures, houses or other real estate}
etc.). There are two types of traders - $(+)$-particles correspond
to \textbf{sellers} and $(-)$-particles to \textbf{buyers, $x_{i}$
}are subjective prices comfortable for the trader $i$. Collision
between particles corresponds to transaction, after this both leave
the market. In more general cases it will be possible that they do
nor leave the market (see below).

We consider here a particular case when for some constant velocities
$v_{\pm}$ and for any $t$ 
\[
f_{\pm}(x,v,t)=\rho_{\pm}(x,t)\delta(v-v_{\pm})
\]
\inputencoding{koi8-r}\foreignlanguage{russian}{For this to hold at
any time $t$ it is sufficient to demand that this holds for $t=0$.
\emph{Initial conditions} are defined by the initial densities $\rho_{\pm}(r,0)$.
The velocities $v_{\pm}$ can be interpreted as averaged velocities
for sellers and buyers correspondingly. }

\paragraph{Two phases - fluid dynamics}

\inputencoding{latin9}It can occur that under some scaling the defined
particle dynamics tends to some kind of\textbf{ continuous (fluid)
picture,} see \cite{m15}, but we shall not pursue this way here.
Instead, we consider continuous densities of $(+)$-masses and $(-)$-masses
and shall define their dynamics directly. We assume that at each time
$t$ there exists point $b(t)$ - boundary between phases. There are
two phases with initial densities $\rho_{+}(r,0),\rho_{-}(r,0)$ where
\[
r=r(t)=|x-b(t)|=\pm(x-b(t))
\]
correspondingly. Phases move with velocities $v_{\pm}$correspondingly.
Collision of plus and minus masses (at the point $b(t)$) leads to
their cancellation in equal amount. There is more realistic possibility
- to make the cancellation proportional to the current price, but
we do not consider this possibility \inputencoding{koi8-r}\foreignlanguage{russian}{here. }

\inputencoding{latin9}We obtain equations for the triple $(b(t),\rho_{+}(r,t),\rho_{-}(r,t))$
similarly to the way how the equations of continuum mechanuics are
derived in the textbooks, that is using conservation laws. Here there
is only one - mass conservation law.

First of all, obtain the equation for the boundary. Assume $b(t)$
smooth and put $\beta=\frac{db(t)}{dt}$. Then for time $dt$ the
amount of positive mass, reaching the boundary will be 
\[
M_{+}(\beta,t)dt=\int_{v-\beta<0}\int_{r<(-v+\beta)dt}f_{+}(r,v,t)dv+o(dt)=dt\int_{v-\beta<0}f_{+}(0,v,t)(-v+\beta)dv+o(dt)
\]
In fact, income and outcome give the contribution $o(dt)$. Similarly
for negative mass 
\[
M_{-}(\beta,t)dt=\int_{v-\beta>0}\int_{r<(v-\beta)dt}f_{-}(r,v,t)dv+o(dt)=dt\int_{v-\beta>0}f_{-}(0,v,t)(v-\beta)dv+o(dt)
\]

\begin{lemma}

For any $t$ there exists unique $\beta=\beta(t)$ such that 
\begin{equation}
M_{+}(\beta,t)=M_{-}(\beta,t)\label{boundary}
\end{equation}

\end{lemma}

In fact, consider the equation with respect to $\beta$ 
\[
\int_{v-\beta<0}f_{+}(0,v,t)(-v+\beta)dv=\int_{v-\beta>0}f_{-}(0,v,t)(v-\beta)dv
\]
Then if $\beta$ increases,  the right-hand side decreases and the
left-hand side increases.

We can rewrite the equation (\ref{boundary}) in our case 
\begin{equation}
\rho_{+}(0,t)(-v_{+}+\beta(t))=\rho_{-}(0,t)(v_{-}-\beta(t))\label{main_1}
\end{equation}
from where we can get $\beta(t)$ 
\begin{equation}
\beta(t)=\frac{\rho_{+}(0,t)v_{+}+\rho_{-}(0,t)v_{-}}{\rho_{+}(0,t)+\rho_{-}(0,t)}\label{main_1_beta}
\end{equation}
Now we should write the equations for the densities. For $\rho_{+}(r,t)$
we get 
\[
\rho_{+}(r,t+\Delta t)=\rho_{+}(r-(v_{+}-\beta(t))\Delta t,t)-\mu_{+}(r,t)\rho_{+}(r,t)\Delta t+\lambda_{+}(r,t)\Delta t+o(\Delta t)=
\]
\[
=\rho_{+}(r,t)-(v_{+}-\beta(t))\frac{\partial\rho_{+}(r,t)}{\partial r}\Delta t-\mu_{+}(r,t)\rho_{+}(r,t)\Delta t+\lambda_{+}(r,t)\Delta t+o(\Delta t)
\]
In the limit $\Delta t\to0$ 
\begin{equation}
\frac{\partial\rho_{+}(r,t)}{\partial t}=-(v_{+}-\beta(t))\frac{\partial\rho_{+}(r,t)}{\partial r}-\mu_{+}(r,t)\rho_{+}(r,t)+\lambda_{+}(r,t)\label{main_plus}
\end{equation}
Similarly $\rho_{-}(r,t)$: 
\[
\rho_{-}(r,t+\Delta t)=\rho_{-}(r+(v_{-}-\beta(t))\Delta t,t)-\mu_{-}(r,t)\rho_{+}(r,t)\Delta t+\lambda_{-}(r,t)\Delta t+o(\Delta t)
\]
\begin{equation}
\frac{\partial\rho_{-}(r,t)}{\partial t}=(v_{-}-\beta(t))\frac{\partial\rho_{-}(r,t)}{\partial r}-\mu_{-}(r,t)\rho_{-}(r,t)+\lambda_{-}(r,t)\label{main_minus}
\end{equation}

It would be nice to prove accurately that the solution of equations
(\ref{main_1},\ref{main_plus},\ref{main_minus}) exists for any
$t\geq0$ and is unique, but we did not try to do this.

\paragraph*{Fixed points and stationary points}

Assume that the functions $\lambda_{\pm}(r)=\lambda_{\pm}(r,t)$ and
$\mu_{\pm}(r)=\mu_{\pm}(r,t)$ do not depend on $t$ (remind that
they were assumed to have compact support). Denote 
\[
\gamma_{cr}^{(+)}=-v_{+}^{-1}\int_{0}^{\infty}\lambda_{+}(x)\exp\left(\frac{1}{v_{+}}\int_{0}^{x}\mu_{+}(y)dy\right)dx,\,\gamma_{cr}^{(-)}=v_{-}^{-1}\int_{0}^{\infty}\lambda_{-}(x)\exp\left(-\frac{1}{v_{-}}\int_{0}^{x}\mu_{-}(y)dy\right)dx
\]
and 
\[
\gamma_{cr}=\max\left(\gamma_{cr}^{(+)},\frac{v_{-}\gamma_{cr}^{(-)}}{-v_{+}}\right)
\]

We define the fixed point of our dynamics by the conditions: $\beta(t)=0$
and $\rho_{\pm}(r,t)$ do not depend on time. Alternatively the fixed
points are defined as any solutions of the stationary version 
\begin{equation}
\rho_{+}(0)v_{+}+\rho_{-}(0)v_{-}=0\label{fixed_11}
\end{equation}
\begin{equation}
-v_{+}\frac{\partial\rho_{+}(r)}{\partial r}-\mu_{+}(r)\rho_{+}(r)+\lambda_{+}(r)=0\label{fixed1plus}
\end{equation}
\begin{equation}
v_{-}\frac{\partial\rho_{-}(r)}{\partial r}-\mu_{-}(r)\rho_{-}(r)+\lambda_{-}(r)=0\label{fixed_1minus}
\end{equation}
of the system (\ref{main_1},\ref{main_plus},\ref{main_minus}).
We will prove that there exists a family of fixed points depending
on a real parameter.

Similarly, we call stationary point any solution of the system of
equations (\ref{main_1},\ref{main_plus},\ref{main_minus}), where
$\beta=\beta(t)$ and the densities do not depend on $t$. We shall
prove that there is a family of stationary points depending on two
real parameters.

We say that a fixed (or stationary) point has finite mass if 
\[
\int_{0}^{\infty}\rho_{\pm}(r)dr<\infty
\]

\begin{theorem}\label{th1}

Let the parameters $\lambda_{\pm}(r),\mu_{\pm}(r)$ and $v_{\pm}$be
fixed. Then 
\begin{enumerate}
\item For any value of the parameter $\gamma_{+}=\rho_{+}(0)$ there is
at most one fixed point. For $\gamma_{+}<\gamma_{cr}$ there is no
any fixed point. For $\gamma_{+}\geq\gamma_{cr}$ there exists exactly
one fixed point defined by\inputencoding{koi8-r}\foreignlanguage{russian}{
\begin{eqnarray}
\rho_{+}(r) & = & \exp\left(-\frac{1}{v_{+}}\int_{0}^{r}\mu_{+}(x)dx\right)\left(\rho_{+}(0)+v_{+}^{-1}\int_{0}^{r}\lambda_{+}(x)\exp\left(\frac{1}{v_{+}}\int_{0}^{x}\mu_{+}(y)dy\right)dx\right)\label{rho_plus}\\
\rho_{-}(r) & = & v_{-}^{-1}\exp\left(\frac{1}{v_{-}}\int_{0}^{r}\mu_{-}(x)dx\right)\left(-v_{+}\rho_{+}(0)-\int_{0}^{r}\lambda_{-}(x)\exp\left(-\frac{1}{v_{-}}\int_{0}^{x}\mu_{-}(y)dy\right)dx\right)\label{rho_minus}
\end{eqnarray}
}\inputencoding{latin9} 
\item The fixed point has finite mass if $\gamma_{cr}^{(+)}=\gamma_{cr}^{(-)}$ 
\item For any $\gamma_{+},\gamma_{-}$such that \inputencoding{koi8-r}\foreignlanguage{russian}{ 
\[
\gamma_{+}=\rho_{+}(0)\geq\gamma_{cr}^{(+)},\;\gamma_{-}=\rho_{-}(0)\geq\gamma_{cr}^{(-)}
\]
there is exactly one stationary point. Then the densities are defined
by formulas (\ref{rho_plus},\ref{rho_minus}) and the boundary velocity
is 
\[
\beta=\frac{\rho_{+}(0)v_{+}+\rho_{-}(0)v_{-}}{\rho_{+}(0)+\rho_{-}(0)}
\]
}\inputencoding{latin9} 
\item Stationary point has finite mass iff $\gamma_{+}=\gamma_{cr}^{(+)},\;\gamma_{-}=\gamma_{cr}^{(-)}$. 
\end{enumerate}
\end{theorem}

Proof. Solving equations (\ref{main_plus},\ref{main_minus}) we get
for any $r>0$ equations (\ref{rho_plus}) and (\ref{rho_minus}).
Note that, by equations (\ref{rho_minus}) and (\ref{rho_plus}),
densities $\rho_{-}(r)$, $\rho_{+}(r)$ are positive iff $\gamma_{+}\geq\gamma_{cr}^{(+)}$,
$\gamma_{-}\geq\gamma_{cr}^{(-)}$. Taking into account equation (\ref{fixed_11})
we get the first asserion of the theorem.

For the stationary points the densities are again defined by equations
(\ref{rho_plus}) and (\ref{rho_minus}). We have two conditions for
them to be non-negative. Then the boundary will move with constant
velocity defined from equation (\ref{main_1_beta}).

\paragraph{More complicated one market model }

Note that collision of masses of two phases create total annihilation
flow 
\[
\nu(t)=(v_{-}-\beta(t))\rho_{-}(0,t)=-(v_{+}-\beta(t))\rho_{+}(0,t)
\]
of the disappearing $(\pm)$-particles. Here we assume that a part
of annihilating particles does not disappear but can transform to
particles of the other phase jumping from the collision point $0$
to some point $r$. On the language of continuous media this means
that there are output flows of mass $\nu(+,-,r,t)$ and $\nu(-,+,r,t)$
such that 
\[
\int_{0}^{\infty}p(+,-,r,t)dr\leq1,\,\int_{0}^{\infty}p(-,+,r,t)dr\leq1
\]
where 
\[
p(+,-,r,t)=\frac{\nu(+,-,r,t)}{\nu(t)},\, p(-,+,r,t)=\frac{\nu(-,+,r,t)}{\nu(t)}
\]
For such model we have the system of three equations 
\begin{eqnarray}
\beta(t) & = & \frac{v_{-}\rho_{-}(0,t)+v_{+}\rho_{+}(0,t)}{\rho_{-}(0,t)+\rho_{+}(0,t)}\nonumber \\
\frac{\partial\rho_{+}(r,t)}{\partial t} & = & -(v_{+}-\beta(t))\frac{\partial\rho_{+}(r,t)}{\partial r}-\mu_{+}(r,t)\rho_{+}(r,t)+\lambda_{+}(r,t)+(v_{-}-\beta(t))\rho_{-}(0,t)p(-,+,r,t)\label{eq:nsts}\\
\frac{\partial\rho_{-}(r,t)}{\partial t} & = & (v_{-}-\beta(t))\frac{\partial\rho_{-}(r,t)}{\partial r}-\mu_{-}(r,t)\rho_{-}(r,t)+\lambda_{-}(r,t)-(v_{+}-\beta(t))\rho_{+}(0,t)p(+,-,r,t)\nonumber 
\end{eqnarray}
We again assume that the functions $\mu_{\pm}(r,t),\lambda_{\pm}(r,t),p(-,+,r,t),p(+,-,r,t)$
do not depend on $t$ and have compact support. Introduce the functions

\[
F_{+}(x)=-\frac{1}{v_{+}}\int_{0}^{x}\mu_{+}(y)dy,\; F_{-}(x)=\frac{1}{v_{-}}\int_{0}^{x}\mu_{-}(y)dy
\]
Denote

\[
\alpha_{-+}=\int_{0}^{\infty}p(-,+,x)\exp\left(-F_{+}(x)\right)dx,\;\alpha_{+-}=\int_{0}^{\infty}p(+,-,x)\exp\left(-F_{-}(x)\right)dx
\]
and assume that $\alpha_{-+},\,\alpha_{+-}<1$. Define 
\[
\hat{\gamma}_{cr}=\max\left(\frac{\gamma_{cr}^{(+)}}{1-\alpha_{-+}},\frac{v_{-}\gamma_{cr}^{(-)}}{-v_{+}(1-\alpha_{+-})}\right)
\]
\begin{theorem}

Let the parameters $\lambda_{\pm}(r),\mu_{\pm}(r),$ $p(-,+,r,t),p(+,-,r,t)$
and $v_{\pm}$be given. Then 
\begin{enumerate}
\item For any value of the parameter $\gamma_{+}=\rho_{+}(0)$ there is
at most one fixed point. For $\gamma_{+}<\hat{\gamma}_{cr}$ there
is no any fixed point. For $\gamma_{+}\geq\hat{\gamma}_{cr}$ there
exists exactly one fixed point. It is 
\[
\rho_{+}(r)=\exp\left(F_{+}(r)\right)\left(\rho_{+}(0)+v_{+}^{-1}\int_{0}^{r}\left(\lambda_{+}(x)-v_{+}\rho_{+}(0)p(-,+,x)\right)\exp\left(-F_{+}(x)\right)dx\right)
\]
\[
\rho_{-}(r)=v_{-}^{-1}\exp\left(F_{-}(r)\right)\left(-v_{+}\rho_{+}(0)-\int_{0}^{r}\left(\lambda_{-}(x)-v_{+}\rho_{+}(0)p(+,-,x)\right)\exp\left(-F_{-}(x)\right)dx\right)
\]

\item There is a unique stationary point with finite mass. It is 
\[
\rho_{+}(r)=\exp\left(F_{+}(r)\right)\left(\rho_{+}(0)+v_{+}^{-1}\int_{0}^{r}\left(\lambda_{+}(x)+(v_{-}-\beta)\rho_{-}(0)p(-,+,x)\right)\exp\left(-F_{+}(x)\right)dx\right)
\]
\[
\rho_{-}(r)=\exp\left(F_{-}(r)\right)\left(\rho_{-}(0)-v_{-}^{-1}\int_{0}^{r}\left(\lambda_{-}(x)-(v_{+}-\beta)\rho_{+}(0)p(+,-,x)\right)\exp\left(-F_{-}(x)\right)dx\right)
\]
and we denote 
\[
\rho_{+}(0)=\frac{-v_{+}\gamma_{cr}^{(+)}}{-v_{+}(1-\alpha_{-+})-\beta\alpha_{-+}}
\]
\[
\rho_{-}(0)=\frac{v_{-}\gamma_{cr}^{(-)}}{v_{-}(1-\alpha_{+-})+\beta\alpha_{+-}}
\]
where $\beta$ is a root (belonging to the interval $(v_{+},v_{-})$)
of \inputencoding{koi8-r}\foreignlanguage{russian}{quadratic equation
(\ref{eq:qu_eq}). It exists and is unique.}\inputencoding{latin9} 
\end{enumerate}
\end{theorem}

Proof. 1. Similarly to the first part of theorem \ref{th1}.

2. As follows from system (\ref{eq:nsts}) the equations for the stationary
points are 
\begin{eqnarray}
0 & = & (v_{-}-\beta)\rho_{-}(0)+(v_{+}-\beta)\rho_{+}(0)\nonumber \\
0 & = & -v_{+}\frac{\partial\rho_{+}(r)}{\partial r}-\mu_{+}(r)\rho_{+}(r)+\lambda_{+}(r)+(v_{-}-\beta)\rho_{-}(0)p(-,+,r)\label{eq:bal-1}\\
0 & = & v_{-}\frac{\partial\rho_{-}(r)}{\partial r}-\mu_{-}(r)\rho_{-}(r)+\lambda_{-}(r)-(v_{+}-\beta)\rho_{+}(0)p(+,-,r)\nonumber 
\end{eqnarray}
Solving these linear first order equations we get

\begin{equation}
\rho_{+}(r)=\exp\left(F_{+}(r)\right)\left(\rho_{+}(0)+v_{+}^{-1}\int_{0}^{r}\left(\lambda_{+}(x)+(v_{-}-\beta)\rho_{-}(0)p(-,+,x)\right)\exp\left(-F_{+}(x)\right)dx\right)\label{eq:pl-1}
\end{equation}
\begin{equation}
\rho_{-}(r)=\exp\left(F_{-}(r)\right)\left(\rho_{-}(0)-v_{-}^{-1}\int_{0}^{r}\left(\lambda_{-}(x)-(v_{+}-\beta)\rho_{+}(0)p(+,-,x)\right)\exp\left(-F_{-}(x)\right)dx\right)\label{eq:min-1}
\end{equation}
We are looking for a stationary point with finite mass such that 
\begin{equation}
\int_{0}^{\infty}\rho_{\pm}(r)dr<\infty\label{eq:fin}
\end{equation}
Then by (\ref{eq:pl-1}), (\ref{eq:min-1}), (\ref{eq:fin}), (\ref{eq:bal-1}),
a stationary point is uniquely defined by three parameters $\gamma_{\pm}=\rho_{\pm}(0)$,
$\beta$ which satisfy the following equations

\begin{eqnarray}
-v_{+}\gamma_{+} & = & -v_{+}\gamma_{cr}^{(+)}+(v_{-}-\beta)\gamma_{-}\alpha_{-+}\nonumber \\
v_{-}\gamma_{-} & = & v_{-}\gamma_{cr}^{(-)}-(v_{+}-\beta)\gamma_{+}\alpha_{+-}\label{eq:syst}\\
(v_{-}-\beta)\gamma_{-} & = & -(v_{+}-\beta)\gamma_{+}\nonumber 
\end{eqnarray}
where $\beta\in(v_{+},v_{-})$. We show that this system has a unique
solution. Using the third equation of the system, we get from the
first two 
\begin{equation}
\gamma_{+}=\frac{-v_{+}\gamma_{cr}^{(+)}}{-v_{+}(1-\alpha_{-+})-\beta\alpha_{-+}}\label{eq:rpl}
\end{equation}
\begin{equation}
\gamma_{-}=\frac{v_{-}\gamma_{cr}^{(-)}}{v_{-}(1-\alpha_{+-})+\beta\alpha_{+-}}\label{eq:rmin}
\end{equation}
Substituting these expressions to the third one we come to the quadratic
equation with respect to $\beta$: 
\begin{equation}
(\sigma_{+}\alpha_{+-}-\sigma_{-}\alpha_{-+})(-v_{+}+\beta)(v_{-}-\beta)+(\sigma_{-}v_{+}-\sigma_{+}v_{-})\beta-v_{+}v_{-}(\sigma_{-}-\sigma_{+})=0\label{eq:qu_eq}
\end{equation}
where, for shortness, we denote $\sigma_{+}=-v_{+}\gamma_{cr}^{(+)}$,
$\sigma_{-}=v_{-}\gamma_{cr}^{(-)}$.

Consider first the case when $\sigma_{+}\alpha_{+-}-\sigma_{-}\alpha_{-+}\neq0$
. Note that the boundary velocity should satisfy $v_{+}<\beta<v_{-}$.
One can show easily that there is always one root of the equation
in the interval $v_{+}<\beta<v_{-}$. Now one should verify that $\gamma_{+},\gamma_{-}$,
defined by (\ref{eq:rpl}) and (\ref{eq:rmin}) are non-negative.
By (\ref{eq:rpl}) (\ref{eq:rmin}) one of the values $\gamma_{+},\gamma_{-}$
is always positive. Then by the third equation of the system (\ref{eq:syst})
also the other value is positive as $v_{-}-\beta,\,-v_{+}+\beta>0$.
Thus there exists the unique fixed point satisfying (\ref{eq:pl-1}),
(\ref{eq:min-1}), (\ref{eq:rpl}) (\ref{eq:rmin}).

Is $\sigma_{+}\alpha_{+-}-\sigma_{-}\alpha_{-+}=0$, we have a linear
equation with respect to $\beta$, we gives 
\[
\beta=\frac{\sigma_{-}-\sigma_{+}}{\sigma_{-}v_{-}^{-1}-\sigma_{+}v_{+}^{-1}}
\]
and from (\ref{eq:rpl}) (\ref{eq:rmin}) we get $\gamma_{+}=\sigma_{+}v_{+}^{-1}$
and $\gamma_{-}=\sigma_{-}v_{-}^{-1}$. In this case also a stationary
point exists and is unique.

\paragraph{Networks with many markets}

Let us call the previous model an elementary market. A network is
a set $V$ of elementary markets with similar parameters and variables
indexed by $m\in V$ 
\[
v_{\pm,m},\lambda_{\pm,m}(r,t),\mu_{\pm,m}(r,t),\rho_{\pm,m}(r,t),b_{m}(t),\beta_{m}(t)
\]
There are also other parameters interconnecting the markets. Denote
$\nu_{+,m}(t)$ ($\nu_{-,m}(t)$) the total annihilation flow of $(\pm)$-particles
from the market $m$. As they are equal we denote $\nu_{m}(t)=\nu_{+,m}(t)=\nu_{-,m}(t)$.
Let  
\[
\nu_{k,m}(+,+,r,t),\nu_{k,m}(+,-,r,t),\nu_{k,m}(-,+,r,t),\nu_{k,m}(-,+,r,t)
\]
be the parts of these annihilation flows of $(\pm)$-particles, that
after the transaction on the market $m$, become $(\mp)$-particles
on the market $k$ with the coordinate $r$. Denote 
\[
p_{km}(\pm,\pm,r,t)=\frac{\nu_{k,m}(\pm,\pm,r,t)}{\nu_{m}(t)}
\]
We mean that $p_{km}(+,+,r,t)=p_{km}(-,-,r,t)\equiv0$. Then for any
$k$ and $t$ the conditions 
\[
\sum_{m\in V}\int_{0}^{\infty}p_{km}(+,-,r,t)dr\leq1,\,\sum_{m\in V}\int_{0}^{\infty}p_{km}(-,+,r,t)dr\leq1
\]
should hold. Denote by $|V|$ the cardinality of the set $V$. We
have then the following system of $3|V|$ equations: 
\begin{eqnarray*}
\beta_{m}(t) & = & \frac{v_{-,m}\rho_{-,m}(0,t)+v_{+,m}\rho_{+,m}(0,t)}{\rho_{+,m}(0,t)+\rho_{+,m}(0,t)}\\
\frac{\partial\rho_{+,m}(r,t)}{\partial t} & = & -(v_{+,m}-\beta_{m}(t))\frac{\partial\rho_{+,m}(r,t)}{\partial r}-\mu_{+,m}(r,t)\rho_{+,m}(r,t)+\lambda_{+,m}(r,t)\\
 & + & \sum_{k\in V}(v_{-,k}-\beta_{k}(t))\rho_{-,k}(0,t)p_{km}(-,+,r,t)\\
\frac{\partial\rho_{-,m}(r,t)}{\partial t} & = & (v_{-,m}-\beta_{m}(t))\frac{\partial\rho_{-,m}(r,t)}{\partial r}-\mu_{-,m}(r,t)\rho_{-,m}(r,t)+\lambda_{-,m}(r,t)\\
 & - & \sum_{k\in V}(v_{+,k}-\beta_{k}(t))\rho_{+,k}(0,t)p_{km}(+,-,r,t)
\end{eqnarray*}

\subparagraph*{Fixed points}

Again we assume $\lambda_{\pm,m}(r,t),\mu_{\pm,m}(r,t),p_{km}(\pm,\pm,r,t)$
do not depend on $t$ and have a compact support. Put 
\[
F_{+}^{(m)}(x)=-v_{+,m}^{-1}\int_{0}^{x}\mu_{+,m}(y)dy,\; F_{-}^{(m)}(x)=v_{-,m}^{-1}\int_{0}^{x}\mu_{-,m}(y)dy
\]
\begin{equation}
\hat{\lambda}_{+,m}=\int_{0}^{\infty}\lambda_{+,m}(x)\exp\left(-F_{+}^{(m)}(x)\right)dx,\;\hat{\lambda}_{-,m}=\int_{0}^{\infty}\lambda_{-,m}(x)\exp\left(-F_{-}^{(m)}(x)\right)dx\label{eq:con}
\end{equation}
\[
\alpha_{km}(-,+)=\int_{0}^{\infty}p_{km}(-,+,x)\exp\left(-F_{+}^{(m)}(x)\right)dx,\;\alpha_{km}(+,-)=\int_{0}^{\infty}p_{km}(+,-,x)\exp\left(-F_{-}^{(m)}(x)\right)dx
\]
for $k,m\in V$.

Define matrices $A_{-+},$ $A_{+-}$ with elements $\alpha_{km}(-,+)$
$\alpha_{km}(+,-)$, where $k,m\in V$, and assume, that they have
the following property: 
\begin{equation}
\forall\, k\;\sum_{m\in V}\alpha_{km}(\pm,\pm)\leq1,\;\exists\: k_{0}\;\sum_{m\in V}\alpha_{km}(\pm,\pm)<1\label{eq:subst}
\end{equation}
For two vectors $a=(a_{i})$ and $b=(b_{i})$ we shall write $a\geq b$$(a>b)$
if $a_{i}\geq b_{i}$ $(a_{i}>b_{i})$ for all coordinates. Consider
the following system of inequalities with respect $\overline{s}$
\begin{equation}
\overline{s}(E-A_{-+})\geq\overline{\lambda}_{+},\;\overline{s}(E-A_{+-})\geq\overline{\lambda}_{-}\label{eq:ineqss}
\end{equation}
where $E$ is the identity matrix and $\overline{\lambda}_{\pm}$
are vectors with coordinates $\hat{\lambda}_{\pm,m}$ defined by (\ref{eq:con}).
We say that this system has a positive solution if there is vector
$\overline{s}$ with positive coordinates satisfying both inequalities
in (\ref{eq:ineqss}). Generally, this system may not have a positive
solution. If one of the matrices $A_{-+}$, $A_{+-}$ is diagonal
or zero the set of positive solutions is nonempty.

\begin{theorem}Each solution $\overline{s}=(s_{m},\, m\in V)>0$
of the system (\ref{eq:ineqss}) uniquely defines the fixed point
as follows: 
\[
\rho_{+,m}(r)=-v_{+,m}^{-1}\exp\left(F_{+}^{(m)}(r)\right)\left(s_{m}-\int_{0}^{r}\left(\lambda_{+,m}(x)+\sum_{k\in V}s_{k}p_{km}(-,+,x)\right)\exp\left(-F_{+}^{(m)}(x)\right)dx\right)
\]
\[
\rho_{-,m}(r)=v_{-,m}^{-1}\exp\left(F_{-}^{(m)}(r)\right)\left(s_{m}-\int_{0}^{r}\left(\lambda_{-,m}(x)+\sum_{k\in V}s_{k}p_{km}(+,-,x)\right)\exp\left(-F_{-}^{(m)}(x)\right)dx\right)
\]
If the set of positive solutions of system (\ref{eq:ineqss}) is empty
there is no any fixed point.

\end{theorem}

Proof. The fixed points satisfy the system consisting of $3|V|$ equation:
\begin{eqnarray}
0 & = & v_{-,m}\rho_{-,m}(0,t)+v_{+,m}\rho_{+,m}(0,t)\nonumber \\
0 & = & -v_{+,m}\frac{\partial\rho_{+,m}(r)}{\partial r}-\mu_{+,m}(r)\rho_{+,m}(r)+\lambda_{+,m}(r)+\sum_{k\in V}v_{-,k}\rho_{-,k}(0)p_{km}(-,+,r)\label{eq:bal}\\
0 & = & v_{-,m}\frac{\partial\rho_{-,m}(r)}{\partial r}-\mu_{-,m}(r)\rho_{-,m}(r)+\lambda_{-,m}(r)-\sum_{k\in V}v_{+,k}\rho_{+,k}(0)p_{km}(+,-,r)\nonumber 
\end{eqnarray}
Solving first order linear differential equations we get 
\begin{equation}
\rho_{+,m}(r)=\exp\left(F_{+}^{(m)}(r)\right)\left(\rho_{+,m}(0)+v_{+,m}^{-1}\int_{0}^{r}\left(\lambda_{+,m}(x)+\sum_{k\in V}v_{-,k}\rho_{-,k}(0)p_{km}(-,+,x)\right)\exp\left(-F_{+}^{(m)}(x)\right)dx\right)\label{eq:pl}
\end{equation}
\begin{equation}
\rho_{-,m}(r)=\exp\left(F_{-}^{(m)}(r)\right)\left(\rho_{-,m}(0)-v_{-,m}^{-1}\int_{0}^{r}\left(\lambda_{-,m}(x)-\sum_{k\in V}v_{+,k}\rho_{+,k}(0)p_{km}(+,-,x)\right)\exp\left(-F_{-}^{(m)}(x)\right)dx\right)\label{eq:min}
\end{equation}
for $m\in V$.

Using equations $0=v_{-,m}\rho_{-,m}(0)+v_{+,m}\rho_{+,m}(0)$, we
conclude that solutions (\ref{eq:pl}), (\ref{eq:min}) are uniquely
defined by parameters $s_{m}=-v_{+,m}\rho_{+,m}(0)$, $m\in V$, and
one can write 
\begin{equation}
\rho_{+,m}(r)=-v_{+,m}^{-1}\exp\left(F_{+}^{(m)}(r)\right)\left(s_{m}-\int_{0}^{r}\left(\lambda_{+,m}(x)+\sum_{k\in V}s_{k}p_{km}(-,+,x)\right)\exp\left(-F_{+}^{(m)}(x)\right)dx\right)\label{eq:pl_g}
\end{equation}
\begin{equation}
\rho_{-,m}(r)=v_{-,m}^{-1}\exp\left(F_{-}^{(m)}(r)\right)\left(s_{m}-\int_{0}^{r}\left(\lambda_{-,m}(x)+\sum_{k\in V}s_{k}p_{km}(+,-,x)\right)\exp\left(-F_{-}^{(m)}(x)\right)dx\right)\label{eq:min_g}
\end{equation}

Whereas the densities (\ref{eq:pl_g}), (\ref{eq:min_g}) are nonnegative
for all $r\geq0$ the following conditions must be satisfied 
\begin{equation}
s_{m}\geq\hat{\lambda}_{+,m}+\sum_{k\in V}s_{k}\alpha_{km}(-,+)\label{eq:c1}
\end{equation}
\begin{equation}
s_{m}\geq\hat{\lambda}_{-,m}+\sum_{k\in V}s_{k}\alpha_{km}(+,-)\label{eq:c2}
\end{equation}
for all $m\in V$. These inequalities are equvalent to system (\ref{eq:ineqss}).

So the fixed points exist iff there exist positive solutions of system
(\ref{eq:ineqss}).


\begin{thebibliography}{10}
\bibitem{ros} A. Rosenfeld. Picture Languages. Acad. Press. 1979.

\bibitem{hab} A. Habel. Hyperedge Replacement: Grammars and Languages.
Lecture Notes in Computer Science, v. 643, 1992. Springer Verlag.

\bibitem{pal} E.M.Palmer. Graphical Evolution. Wiley. 1985.

\bibitem{m3} V.A.Malyshev. Interacting Strings of Symbols. Russian
Math. Surveys, 1997, v.52, No. 2, 59-86.

\bibitem{m1} V.A.Malyshev. Random Grammars. Russian Math. Reviews,
1998, No. 2, pp.

\bibitem{bo} B. Bollobas. Random Graphs.1985. Academic Press.

\bibitem{erre} P. Erdos, A. Renyi. On the evolution of random graphs.
Bull. Inst. Int. Statist. Tokyo, 1961, v. 38,343-347.

\bibitem{kar} M. Karonski. Random Graphs. In Handbook of Combinatorics,
vol. 1 (Eds. R. Graham,M. Grotschel, L. Lovasz). 1995, Elsevier.

\bibitem{Liggett} Th. Liggett. Interacting Particle Systems. 1985.
Springer.

\bibitem{m4}V. A. Malyshev, R. A. Minlos. Gibbs random fields. 1991.
Kluwer.

\bibitem{m5}V. Malyshev, A. Yakovlev. Condensation in Large Closed
Jackson Networks. Annals of Applied Probability, 1996, v.6, No. 1,
pp. 92-115.

\bibitem{m6}A. Zamyatin, V. Malyshev. Introduction to stochastic
models of transportation flows. In the book ``Introduction to mathematical
modelling of transportation flows'', 2010, Moscow, pp. 247-287.

\bibitem{m7}V. Malyshev. Networks and dynamical systems Adv. Appl.
Prob., 1993, v. 25, 140-175.

\bibitem{m8}V. Malyshev. Microscopic Models of Chemical Thermodynamics,
2005, J. Stat,. Physics, 119, No. 5/6, pp. 997-1026.

\bibitem{m9}V. Malyshev, A. Manita, A. Zamyatin. Homeostasis phenomena
in chemical reaction networks. Probability theory and applications,
2006, v. 51, pp. 793-802.

\bibitem{m10}V. Malyshev. Macrodimension is an invariant of local
dynamics. Probability theory and applications, 2000, 45, No.2, 368-374.

\bibitem{m11}V. Malyshev. Gibbs and quantum discrete spaces. Russian
Math. Surveys, 2001, v. 56, No. 5, pp. 117-172.

\bibitem{Janson}S. Janson, T. Luczak, A. Ruchinski. Random Graphs.
2000. Wiley.

\bibitem{Kolchin}V. Kolchin. Random Graphs. 2004. Moscow.

\bibitem{Sattoras}R. Pasto-Sattoras, M. Rubi, A. Diaz-Guilera (Eds.)
Statistical Mechanics of Complex Networks. Lecture Notes in Physics.
2003, Springer.

\bibitem{m12}V. Malyshev. Random graphs and Graph Grammars. Discrete
Mathematics and its applications, 1998, v. 8. No. 3, 247-262.

\bibitem{m13}V. Malyshev. Random Infinite Spin Graph Evolution. In
``On Dobrushin's way. From Probability Theory to Statistical Physics
AMS Publications, v. 198, 2000, pp. 157-167.

\bibitem{m15}V. A. Malyshev, A.D. Manita. Dynamics of phase boundary
with particle annihilation. Markov Processes and Related Fields, 2009,
v. 15, No. 4, 575-584.

\bibitem{m16}V. A. Malyshev, A.D. Manita, A. A. Zamyatin. Explicit
asymptotic velocity of the boundary between particles and antiparticles.
ISRN Mathematical Physics, 2012.

\bibitem{m17}V. Malyshev. Combinatorics and probability of maps In
\textquotedbl{}Asymptotic Combinatorics with Applications to Mathematical
Physics\textquotedbl{}, Kluwer, 2002, pp. 71-95.

\bibitem{m18}V. Malyshev. Quantun Evolution of Words. Theoretical
Computer Science, 2002, v. 273, pp. 263-269.

\bibitem[27]{ReedSimon}M. Reed, B. Simon. Methods of Mathematical
Physics, v. 3, 1979. Academic Press.\end{thebibliography}
\end{document}